# Pressure-induced inverse order-disorder transition in double perovskites

Zheng Deng[1,2*], Chang-Jong Kang[3*], Mark Croft[3], Wenmin Li[1], Xi Shen[1], Jianfa Zhao[1], Richeng Yu[1], Changqing Jin[1*], Gabi Kotliar[3], Sizhan Liu[4], T. A. Tyson[4], Ryan Tappero[5], Martha Greenblatt[2*]

[1]Beijing National Laboratory for Condensed Matter Physics, and Institute of Physics, Chinese Academy of Sciences, Beijing, 100190, China.
[2]Department of Chemistry and Chemical Biology, Rutgers, the State University of New Jersey, 610 Taylor Road, Piscataway, NJ 08854, USA
[3]Department of Physics and Astronomy, Rutgers, the State University of New Jersey, 136 Frelinghuysen Road, Piscataway, NJ 08854, USA.
[4]Department of Physics, New Jersey Institute of Technology, Newark, NJ 07102, USA.
[5]Brookhaven National Laboratory, Photon Sciences Division, Upton, New York 11973
Corresponding authors: Z.D. (email: dengzheng@iphy.ac.cn), C.-J.K. (email: ck620@physics.rutgers.edu), C.J. (jin@iphy.ac.cn) or M.G. (email: greenbla@chem.rutgers.edu)

**Abstract**

Given the consensus that pressure improves cation order in most of known materials, a discovery of pressure-induced disorder could require reconsideration of order-disorder transition in solid state physics/chemistry and geophysics. Double perovskites $Y_2CoIrO_6$ and $Y_2CoRuO_6$ synthesized at ambient pressure show $B$-site order, while the polymorphs synthesized at 6 and 15 GPa are partially-ordered and disordered respectively. With the decrease of ordering degrees, the lattices are shrunken and the crystal structures alter from monoclinic to orthorhombic symmetry. Correspondingly, long-range ferrimagnetic order in the $B$-site ordered phases are gradually overwhelmed by $B$-site disorder. Theoretical calculations suggest that unusual unit cell compressions under external pressures unexpectedly stabilize the disordered phases of $Y_2CoIrO_6$ and $Y_2CoRuO_6$.

**Introduction**

As a fundamental thermodynamic variable, pressure is experimentally controllable to tune and alter structural, chemical, physical, and mechanical properties. It can also extend new phases by altering crystal structures or atomic orders of an existing material. The discovery of post-perovskite phase under extremely high-pressure successfully explained the origin of the D" seismic discontinuity in lower mantle[1, 2]. Via modifying long-range hydrogen ordering, pressure can modify the ferroelectric state of the potassium dihydrogen phosphate[3, 4].

Order-disorder effects in materials are of fundamental interest to solid state sciences[5-7]. Polymorphs with identical chemical formula, but different ordering degrees, can have altered crystal symmetry and hence distinct properties. Order-disorder transitions have been experimentally and theoretically studied in alloys and mineralogical systems[6-12]. In addition to temperature, pressure has also been applied to modify atomic ordering, owing to the development of high-pressure techniques[5, 13-15]. Although pressure generally improves cation order, particularly in most rock-forming minerals, obtaining insight of pressure-controlled order-disorder transition could generate new techniques to improve performances of materials, and build more rational models of transport properties of minerals that constitute the deep interiors of our planet, as the post-perovskite demonstrated. Double perovskites (DP) with formula $A_2B'B''O_6$ containing exactly 50% $B'$ and 50% $B''$ cations at the $B$-site, can form $B$-site ordered, partially ordered or disordered polymorphs[16]. Hence this family provides ideal models to investigate order-disorder effects which have attracted extensive attention of researchers[17-21].

Generally large differences in oxidation states and cation radii of *B*' and *B*'' tends to enhance cation order in DPs. In this case the disordering will raise the energy, due to local strain/distortion from neighboring *B*'-*B*'/*B*''-*B*'' and electrostatic repulsion from highly charged *B*''-*B*'' (*B*'' is defined with higher oxidation state than *B*'), thereby stabilizing the *B*-site ordered phase[20]. The disordered phases will also have larger lattice volumes, since the distance of neighboring *B*''-*B*'' will be additionally enlarged owing to strong repulsion[18]. The application of pressure always drives smaller cell volumes and consequently, stabilizes the cation ordered phases over the disordered phases. In multiple experiments, synthesis pressure has been used to improve *B*-site order of the ambient-phases[22-24]. For example, $Sr_2FeReO_6$ obtained under ambient pressure has $s \approx 0.7$ while the high-pressure-synthesized sample is nearly completely ordered ($s \approx 1$) [25].

This simple mechanism for pressure-induced ordering in DPs works well until one finds higher synthesis pressure decreasing *B*-site order in $Y_2CoIrO_6$ (YCIO) and $Y_2CoRuO_6$ (YCRO) as shown in this work. The former is a new DP while the ordered phase of the latter was recently reported[26]. In YCIO the ordering degree decreases form $s = 0.854$ for ambient pressure synthesized sample (YCIO-0G) to $s = 0.454$ for sample synthesized at 6 GPa (YCIO-6G), and finally to $s = 0$ for sample synthesized at 15 GPa (YCIO-15G). For YCRO $s$ changes form 0.84 of ambient pressure sample (YCRO-0G) to 0 of 15 GPa one (YCRO-15G). Correspondingly, substantial decreases of unit cell volume of YCIO and YCRO are experimentally observed. For both compounds, the structure symmetries and magnetic properties also synchronously change with the loss of *B*-site ordering.

**Results**

**Crystal Structures.** The high scattering contrast to X-ray between Co and Ir/Ru allows us to quantitatively analyze the *B*-site ordering degrees. The YCIO-0G and YCRO-0G samples show nearly complete rock-salt ordered *B*'- and *B*''-sublattices, while YCIO-6G shows *B*-site partial-order, whereas YCIO-15G and YCRO-15G show complete *B*-site disorder. Figure 1(a) displays room-temperature PXD patterns for YCIO-0G, YCIO-6G and YCIO-15G. The top panel shows PXD pattern for a typical *B*-site rock-salt ordered DP with monoclinic $P2_1/n$ symmetry. In this model, *B*'(Co)- and *B*''(Ir)-site cations occupy two separate crystallographic sites, 2*c* (1/2, 0, 1/2) and 2*d* (1/2, 0, 0), respectively. The main difference among the top three panels is the intensity of the peaks around 19.5º which are from the ordered *B*-site reflections ((011) peaks)[27]. In the insets of Figure 1(a), the normalized (011) peak is pronounced for YCIO-0G, while it decays dramatically for YCIO-6G and vanishes for YCIO-15G, indicating the loss of *B*-site order. Consequently, YCIO-0G and YCIO-6G are refined with $P2_1/n$ space group by Rietveld analysis. The obtained parameters are tabulated in Table 1 and 4S. Figure 1 (b) demonstrates that YCIO-0G is an isostructural analogue of La$_2$CoIrO$_6$, but with stronger CoO$_6$-IrO$_6$ octahedral tilting[28]. For YCIO-0G, the bond valence sum (BVS) is 2.12 for Co and 4.09 for Ir, consistent with Co$^{2+}$ and Ir$^{4+}$ evidenced by XANES results (Figure S5-S6).

The major difference between YCIO-0G and YCIO-6G is the decreasing *B*-site order (Figure 1 (c)) and smaller cell volume with higher pressure ($s_{YCIO-0G}$ = 0.854, $s_{YCIO-6G}$ = 0.454; $V_{YCIO-0G}$ = 227.053 Å$^3$, $V_{YCIO-6G}$ = 226.268 Å$^3$). The cell volume of YCIO-15G decreases to 225.696 Å$^3$. Initial $P2_1/n$ refinement of YCIO-15G yielded $s$ = 0.038, a value extremely close to complete *B*-site disorder. Therefore, orthorhombic symmetry with space group *Pbnm* for *B*-site disordered DP was used[20]. In this model, the 2*c* and 2*d* sites in $P2_1/n$ merge into one crystallographic site, 4a (0, 0.5, 0), which exactly contain 50% Co and 50% Ir for the case of YCIO (Figure 1 (d)). Thus YCIO-15G possesses the lowest *B*-site order ($s$ = 0) and smallest lattice volume. The space groups of *B*-site ordered YCIO-0G and disordered YCIO-15G are further confirmed by TEM: along the [100] zone-axis the (0 2*k*+1 *l*)-serial spots in the selected area electron diffraction patterns are present for YCIO-0G, but extinct for YCIO-15G (Figure S7). In short, we obtained three phases of YCIO with three degrees of *B*-site ordering: a nearly completely ordered YCIO-0G, a partially-ordered YCIO-6G and a completely disordered YCIO-15G. It is noteworthy that XANES measurements suggest a slight valence change of $Co^{2+}$ and $Ir^{4+}$ in YCIO-0G to $Co^{2.14+}$ and $Ir^{3.70+}$ in YCIO-15G (Figure S5-S6).

Similar pressure-induced order-disorder transition occurs in YCRO. YCRO-0G is isostructural to YCIO-0G with $s_{YCRO-0G}$ = 0.84, consistent with previous report[26]. In contrast, in the inset of Figure 1(a) the extinct (011) peak on the PXD pattern of YCRO-15G implies complete *B*-site disorder. Rietveld refinement indicates that YCRO-15G is isostructural to YCIO-15G (Table 1 and Figure 1(d)). The lattice volume $V$ = 225.112 Å$^3$ is substantially smaller than that of YCRO-0G (227.198

Å$^3$)[26].

**Magnetic properties.** The degrees of *B*-site order significantly influence the magnetic behavior of both YCIO and YCRO. Similar to previous studies magnetic frustration, which is induced by *B'*-*B''* antisite[22, 25, 29], weakens long-range magnetic ordering in $A_2B'B''O_6$. Figure 2(a) shows a ferromagnetic (FM)-like phase transition (transition temperature $T_C$ ~ 123 K) in YCIO-0G upon cooling down. The divergence between ZFC and FC and the "λ"-shape feature indicate the presence of magnetic frustration, presumably due to minor *B*-site disorder. Nevertheless, the long-range magnetic ordering phase transition is evidenced by a clear "λ"-shape peak in the specific heat versus temperature ($C_P(T)$) curve (Figure S8). The saturated hysteresis loop in Figure 2(d) evidences a robust magnetic order. Above $T_C$, a hyperbola-like inverse susceptibility ($\chi^{-1}(T)$) is a feature of ferrimagnetism (FiM)[26]. The Curie-Weiss law, $1/\chi = (T - \theta)/C$, is used for fitting the data in the high-temperature region (250−400 K), which yields a Curie constant $C$ = 2.84 emu·K/mol/Oe with an effective moment $\mu_{eff}$ = 4.77 $\mu_B$/*f.u.*, and the paramagnetic temperature $\theta$ = -18.6 K. Both $\mu_{eff}$ and $\theta$ are comparable to those of La$_2$CoIrO$_6$[28].

In Figure 2(b) YCIO-6G also shows FM-like transitions at 103 K, a lower temperature than $T_C$ of YCIO-0G. However, the weakened magnetization of YCIO-6G than that of YCIO-0G in either $M(T)$ or $M(H)$ (Figure 2(d)), along with the broader hump around $T_C$ of $C_P(T)$ (Figure S8) indicate strong magnetic frustration and loss of magnetic ordering. We obtain $\mu_{eff}$ = 4.72 $\mu_B$/*f.u.* and $\theta$ = -41.3 K from the Curie-Weiss fit. In Figure 2(c), under $H$ = 0.1 T, the shape peak at 50 K suggests an antiferromagnetic (AFM) transition. However, with increasing $H$, the peak is smeared and the divergence between ZFC and FC is vanishing. In Figure 2(d) the $M(H)$ loop for YCIO-15G is slightly open even at $H \sim 6$ T. In Figure S8 $C_P(T)$ shows no visible sign of phase transition for YCIO-15G. These features demonstrate spin-glass-like short-range magnetic order due to completely disordered $B'B''$-site cations. The $\chi^{-1}(T)$ curve of YCIO-15G is nearly linear above 50 K, a signature of dominating AFM exchange interactions. The obtained Curie-Weiss parameters are $\mu_{eff} = 4.80$ $\mu_B/f.u.$ and $\theta = -82.6$ K. The marginal change of effective moments (4.72-4.80 $\mu_B/f.u.$) among the three phases is coincident with almost unchanged $d$-electron configurations on Co and Ir.

Similarly, the magnetic behaviors of YCRO also change correspondingly with different $B$-site ordering. As shown in Figure 2(e) and 2(f), YCRO-0G exhibits a FiM transition at around 80 K with long-range magnetic ordering[26]. YCRO-15G shows a FM-like transition at 50 K with much stronger magnetic frustration, which is evidenced by an almost closed and unsaturated $M(H)$ loop in Figures 2(e) and 2(f). For YCRO-15G, we obtained Curie-Weiss parameters: $\mu_{eff} = 6.03$ $\mu_B/f.u.$ and $\theta = -328.7$ K. The former is comparable to that of YCIO-0G ($\mu_{eff} = 5.68$ $\mu_B/f.u.$) while that of the latter is over 200% of the ordered phase ($\theta = -162$ K)[26].

**Theoretical analyses.** It is essential to explain the pressure-induced order-disorder phase in this compound. As already noted, usually, pressure tends to increase cation order in a crystal structure. However, YCIO and YCRO show the opposite trend. To explain the unusual behaviors, we applied the recently proposed order-disorder theory to our compounds[19, 30]. The thermodynamic potential, $G \equiv E - TS + PV$, where $E$, $S$, and $P$ are internal energy, entropy and applied pressure, respectively. Then at certain temperature and pressure the difference between ordered and disordered phases is:

$$\Delta G \equiv G_{dis} - G_{ord} = (E_{dis} - TS_{dis} + PV_{dis}) - (E_{ord} - TS_{ord} + PV_{ord}) = \Delta E - T\Delta S + P\Delta V \quad (1)$$

If $\Delta G > 0$, then the ordered phase is favorable, and vice versa. Generally, with increasing disordering one always has $\Delta E > 0$, due to local strain/distortion, and $-T\Delta S < 0$. The third term, $P\Delta V$, becomes crucial for order-disorder transition: it could promote or impede the formation of an ordered phase depending on the sign of $\Delta V$. Namely, pressure assists cation ordering when $\Delta V > 0$ and hinders it, when $\Delta V < 0$.

A rough approximation is that the higher-charged cation ($B''$) determines the average bond length of the disordered phase due to its stronger (than that of the $B'$) bond stiffness (see sections 6 and 7 of Supporting Information). In YCIO and YCRO, $Ir^{4+}$ and $Ru^{4+}$ are significantly smaller than $Co^{2+}$, and thus the Co/Ir-O or Co/Ru-O bond lengths in disordered phases are shorter than 1/2(Co-O + Ir(Ru)-O) in ordered phases. Consequently the disordered phases have smaller unit cells. We further propose a stricter calculation according to the statistical model proposed by Sakhnenko and Ter-Oganessian[19, 30]. Here $G$ can be expressed as parameters $s$ and $a$, where $a$ is the reduced cubic cell lattice parameter corresponding to one $AB'_{1/2}B''_{1/2}O_3$ formula unit cell. From the system of equation, $\frac{\partial G}{\partial a} = 0$, one can determine the equilibrium value of $a$ for a given values of $s$ and $P$. For $A_2^{3+}B'^{2+}B''^{4+}O_6$ at ambient pressure, the equilibrium lattice constant $a$ can be written as

$$a(s) = \frac{18\sqrt{2}l_A + (14+2s^2)l_{B'} + (20-4s^2)l_{B''}}{35-s^2}, \quad (2)$$

where $l_A$, $l_{B'}$, and $l_{B''}$ are unstrained equilibrium bond lengths of A–O, B'–O, and B''–O, respectively. The corresponding values are related to the effective cation radii. Then the equilibrium values $a$ for disordered and ordered phases are (from Eq. (2))

$$a(s=0) = \frac{18\sqrt{2}l_A + 14l_{B'} + 20l_{B''}}{35}, \text{ and } a(s=1) = \frac{9\sqrt{2}l_A + 8l_{B'} + 8l_{B''}}{17}.$$

When $a(s=0) > a(s=1)$, namely, $l_{B''} > \frac{7}{10}l_{B'} + \frac{3\sqrt{2}}{20}l_A$, pressure promotes B-site ordering, otherwise pressure hinders the B-site ordering.

Based on the above mechanism, we plot the general phase diagram for $A_2^{3+}B'^{2+}B''^{4+}O_6$ in Figure 3. The red line determined by the effective cation radius of $Y^{3+}$, i.e. $l_{B''} = \frac{7}{10}l_{B'} + \frac{3\sqrt{2}}{20}l_Y$, divides the diagram into two regions: the top region shows $V_{dis} > V_{ord}$, corresponding to pressure-induced ordering, while pressure enhances disordering at the bottom area. Since the size of cation $Co^{2+}$ (88.5 pm) is substantially larger than that of $Ir^{4+}$ (76.5 pm) and $Ru^{4+}$ (76.0 pm), YCIO and YCRO are located in a region of $\Delta V < 0$, where the disordered phase is more preferred at high pressure.

In Eq. (1), when $P$ is set, the order-disorder transition temperature ($T_{o\text{-}d}$) is therefore obtained from the relation: $\Delta G = 0$. With the oxidation states determined by XANES, which are $Co^{2+}$ and $Ir^{4+}$ for YCIO-0G, $Co^{2.14+}$ and $Ir^{3.7+}$ for YCIO-15G, our calculations show $T_{o\text{-}d} = 2386$ K at $P = 0$ GPa and $T_{o\text{-}d} = 1172$ K at $P = 15$ GPa. We assume the oxidation states of YCRO-15G maintain $Co^{2+}$ and $Ru^{4+}$ in YCRO-0G[26], due to lack of corresponding XANES data. Then $T_{o\text{-}d} = 1389$ K at $P = 0$ GPa and $T_{o\text{-}d} = 619$ K at $P = 15$ GPa are obtained for YCRO. Regardless of the relative quantities, the decreasing tendency of $T_{o\text{-}d}$ upon pressure further supports the finding that pressure induces *B*-site disorder in YCIO and YCRO. It is noteworthy that if the configuration of $Co^{2+}$ and $Ir^{4+}$ is used at $P$ = 15 GPa, the calculated $T_{o-d}$ = 1337 K. This indicates that the slight charge transfer between Co and Ir indeed benefits pressure-induced order-disorder transition in YCIO.

We now turn to investigate the electronic structure of YCIO-0G and YCRO-0G, the *B*-site ordered phases. The experimental crystal structure $P2_1/n$ was adopted for DFT calculations. Figure 4 shows the electronic density of states with FiM ordering. YCIO-0G and YCRO-0G are insulating with band gaps of 0.28 and 1.13 eV, respectively. The electron occupation numbers are: 6.73 and 5.30 for Co-3*d* and Ir-5*d* in YCIO-0G; 6.76 and 4.57 for Co-3*d* and Ru-4*d* in YCRO-0G, respectively. Considering the hybridization with O-2*p* orbital, it suggests $Co^{2+}$ ($3d^7$) and $Ir^{4+}$ ($5d^5$) in YCIO-0G, $Co^{2+}$ ($3d^7$) and $Ru^{4+}$ ($4d^4$) in YCRO-0G[26], consistent with the XANES experiments. In YCIO-0G, we obtain spin moments $\mu_S(Co)$ = 2.69 $\mu_B$, $\mu_S(Ir)$ = -0.26 $\mu_B$, orbital moments $\mu_L(Co)$ = 0.12 $\mu_B$, $\mu_L(Ir)$ = -0.26 $\mu_B$, and total moments $\mu_{tot}(Co)$ = 2.81 $\mu_B$, $\mu_{tot}(Ir)$ = -0.64 $\mu_B$, where the minus sign indicates the opposite direction to the magnetic moment of Co. Therefore, the net magnetic moment is 2.46 $\mu_B$ per formula unit (*f.u.*) by taking into account the induced magnetic moment in the interstitial region. It is much larger than the saturated magnetization ($\mu_{sat}$) of ~ 0.6 $\mu_B$/*f.u.* in Figure 2(d), presumably due to canting effect. In YCRO-0G, the moments are $\mu_S(Co)$ = 2.70 $\mu_B$, $\mu_S(Ru)$ = -1.35 $\mu_B$, $\mu_L(Co)$ = 0.15 $\mu_B$, $\mu_L(Ru)$ = -0.16 $\mu_B$, and $\mu_{tot}(Co)$ = 2.85 $\mu_B$, $\mu_{tot}(Ru)$ = -1.51 $\mu_B$. The net magnetic moment 1.01 $\mu_B$/*f.u.* is also slightly larger than the experimental result of $\mu_{sat}$ ~ 0.8 $\mu_B$/*f.u.*[26].

**Discussion**

We observe a crossover from FiM to spin-glass-like magnetism, despite of uncertain magnetic structures, in YCIO with pressure increasing from 0 to 15 GPa. Considering negligible changes in Co-O-Ir bond angles and nearly unchanged oxidation states of Co and Ir, we infer the *B*-site disordering primary contribute to the change on magnetic behaviors, as proposed in many literatures[22, 31]. It is also consistent with increase of $\theta$ from -18.6 to -82.6 K. The increasing AFM exchange interactions should be primarily from superexchanges of anti-parallel nearest neighbor Co-Co and Ir-Ir yielded by Co-Ir antisite. For YCRO, the substantial increase of paramagnetic temperature from -162 to -328.7 K is also a strong evidence of overwhelming Co-Ru antisite occupation. Thus the considerable changes in magnetic behaviors between ambient and high-pressure synthesized YCIO and YCRO, respectively strongly support evidence of pressure-induced *B*-site disorder in YCIO and YCRO.

High pressure has been widely used to synthesize *B*-site ordered or disordered DPs for decades[32, 33]. However, surprisingly, YCIO and YCRO are the first two DPs, to the best of our knowledge, to show pressure-induced *B*-site disorder. Typically, increasing pressure is used to improve cation order rather than disorder, based on the aforementioned assumption that decreasing cell volume leads to higher ordering.[5] Furthermore, *B*-site ordered $A_2B'^{m+}B''^{n+}O_6$ with n ≥ 5[20], are overwhelmingly more numerous than others. Generally it is difficult to induce disorder by overcoming the electrostatic repulsions between the highly charged *B*''-cations[20]. The only group of DPs with small n and minimum $\Delta n_B$ ($\Delta n_B$ = n-m) is $A_2B'^{2+}B''^{4+}O_6$. Additionally, the slight charge transfer from Co to Ir with increasing $P$, further decreases the valence and bond length difference between $B'$ and $B''$-site cations. As a result, $T_{o-d}$ is further reduced under pressure. Due to the above features, pressure-induced disordering is observed in YCIO and YCRO. It is noteworthy that both ambient-pressure-synthesized $B$-site ordered YCIO and YCRO can be directly converted to $B$-site disordered phases under 15 PGa within a half hour. Thus the rapid order-disorder transformations rule out a mechanism controlled by kinetic effects.

In summary, we synthesized a new completely $B$-site ordered DP, $Y_2CoIrO_6$ at ambient pressure. Surprisingly, when $Y_2CoIrO_6$ is synthesized at high pressure, 6 and 15 GPa, the phases form with smaller unit cell volume, but contrary to expectations, with partial or complete $B$-site disordering, respectively. The $B$-site ordered (0GPa) and partially ordered (6 GPa) phases crystallize in a monoclinic ($P2_1/n$) structure, while the disordered (15 GPa) one forms in orthorhombic ($Pbnm$) symmetry. The increasing Co-Ir antisite generates magnetic frustrations that weaken and finally break long-range ferrimagnetic ordering in the $B$-site ordered $Y_2CoIrO_6$. Nearly identical phenomenon is also found between $B$-site ordered (0GPa) and disordered (15 GPa) $Y_2CoRuO_6$. Both ambient-pressure-synthesized $B$-site ordered YCIO and YCRO can be directly converted rapidly to $B$-site disordered phases under 15 GPa. With DFT calculations and the statistical model of atomic ordering, we infer that this unique pressure-induced order-to-disorder transition in $Y_2CoIrO_6$ and $Y_2CoRuO_6$ are primarily due to the substantially larger effective cation size of $Co^{2+}$ compared to that of $Ir^{4+}(Ru^{4+})$, thereby rendering $\Delta V = V_{dis} - V_{ord} < 0$. Thus $\Delta G = G_{dis} - G_{ord} = \Delta E - T\Delta S$

+ $P\Delta V$ < 0 at high pressure due to the negative $P\Delta V$, so that this unusual volume change ($V_{dis}$ < $V_{ord}$ at high $P$) stabilizes the disordered phase with increasing pressure. Pressure-induced disorder in $Y_2CoIrO_6$ and $Y_2CoRuO_6$ is contrary to traditional theories of order-disorder mechanism and will lead to reconsideration of pressure-effects in solid state sciences.

**Methods**

**Synthesis and laboratory powder X-ray diffractions (PXD).** Polycrystalline YCIO and YCRO samples were synthesized at varying pressures. For the samples synthesized at ambient pressure (YCIO-0G, YCRO-0G), the well ground mixture of $Y_2O_3$ (Alfa Aesar, 99.9%), $Co_3O_4$ (Alfa Aesar, 99.9%), $IrO_2$ (Alfa Aesar, 99.9%) and $RuO_2$ (Alfa Aesar, 99.9%) was pelletized and then calcined at 1500 K for 12 h in air. After intermediate grinding, the recovered material was pressed into pellets again and calcined for another 5 h at 1500 K in air. The high-pressure samples (YCIO-6G, YCIO-15G and YCRO-15G) were synthesized at Institute of Physics CAS. The well ground stoichiometric mixture of $Y_2O_3$, $Co_3O_4$, Co (Alfa Aesar, 99.95%), $IrO_2$ and $RuO_2$ was placed in Pt capsules and then calcined at 1500 ~ 1550 K for 0.5 h under 6 and 15 GPa with a Walker-type multianvil high-pressure apparatus, respectively. The pressures were maintained during the temperature quenching and then decompressed slowly after reactions. It is noteworthy that YCIO-6G, YCIO-15G and YCRO-15G also can be synthesized with YCIO-0G and YCRO-0G samples as precursors under the same reaction conditions. Phase purity and crystal symmetry of all the samples were determined with laboratory PXD. Rietveld refinements were performed using the GSAS software packages[34].

**Energy dispersive analysis (EDX) and transmission electron microscopy (TEM).** All the YCIO and YCRO samples, either ambient pressure or high-pressure made, are dark gray. The observed compositions of the heavy elements (i.e. Y, Co, Ir and Ru) are consistent with their nominal stoichiometry as determined with EDX (Figure S1-S4 and Table S1-S4). Microstructures and crystal symmetries (related with *B*-site ordering degrees) of YCIO-0G and YCIO-15G were studied with a high resolution TEM (JEOL ARM200F).

**X-ray absorption near-edge spectroscopy (XANES).** The Ir and Co XANES experiments on the YCIO samples were performed respectively at the next-generation National Synchrotron Light Source (NSLS-II) beamlines: 7-BM QAS with a Si-111 double crystal monochromator; and 4-BM XFM with a Si-220 double crystal monochromator. The 7-BM measurements performed in both transmission and fluorescent modes with a simultaneous standard. The 4-BM XFM measurements were performed using a defocused beam in the fluorescence mode with a proximately run standard. Standard pre- and post-background subtraction normalization to unity absorption step at the edge were used in the analysis. Many of the standard spectra were collected on NSLS-I on beamline X19A using a Si-111 double crystal monochromator.

**DC magnetization and heat capacity.** For all the samples, DC magnetization was measured with a Quantum Design Superconducting Quantum Interference

Device-Vibrating Sample Magnetometer (SQUID-VSM). Temperature-dependent magnetization ($M(T)$) was measured under both zero field cooling (ZFC) and field cooling (FC) procedures. Magnetic field dependence of magnetization ($M(H)$) was measured within a field range of ±7 T at selected temperatures. Heat capacity measurements on YCIO samples were conducted with a Quantum Design Physical Property Measurement System (PPMS).

**Theoretical calculations.** The density functional theory (DFT) calculations of YCIO and YCRO were performed with the all-electron full-potential linearized augmented plane-wave (FLAPW) method implemented in the WIEN2k code.[35] Generalized gradient approximation (GGA) of Perdew-Burke-Ernzerhof (PBE) was used for the exchange-correlation functional[36]. Due to the heavy elements of Ir and Ru in title compounds, the spin-orbit coupling was taken into account in the second variation method. In order to consider the correlation effect, GGA + U was adopted within the fully localized limit[37, 38]. The on-site Coulomb interaction parameters, U = 7 and 2 eV for Co and Ir in YCIO, while U = 7 and 4 eV for Co and Ru in YCRO, respectively were used. In the DFT calculation, ferrimagnetic order was considered. To calculate the order-disorder phase transition temperatures, we adopted the theory recently suggested by Ter-Oganessian and Sakhnenko[19, 30]. Our calculations explain the order-disorder phase transitions under pressure in YCIO and YCRO.

**Acknowledgments**

C.-J.K., G.K., and M.G. were supported by the U. S. Department of Energy, Office of Science, Basic Energy Science as a part of the Computational Materials Science Program through the Center for Computational Design of Functional Strongly Correlated Materials and Theoretical Spectroscopy. M.G. also acknowledges support of NSF-DMR-1507252 grant. Work at Brookhaven National Laboratory was supported by the DOE BES (DE-SC0012704) on the NSLS-II Beamlines 7BM and 6BM. Works at IOPCAS were supported by MOST (No. 2018YFA03057001 and 2017YFB0405703) & NSF (No. 11974407, 11820101003, 11921004) of China through research projects. Z.D. also acknowledges support of the Youth Innovation Promotion Association of CAS. The authors wish to gratefully acknowledge D. Walker, X.-H. Xu, F.-X. Jiang and J.-G. Cheng for helpful discussions.

Competing interests: The authors declare no competing interests.

**Figures and Table**

**Figure 1**

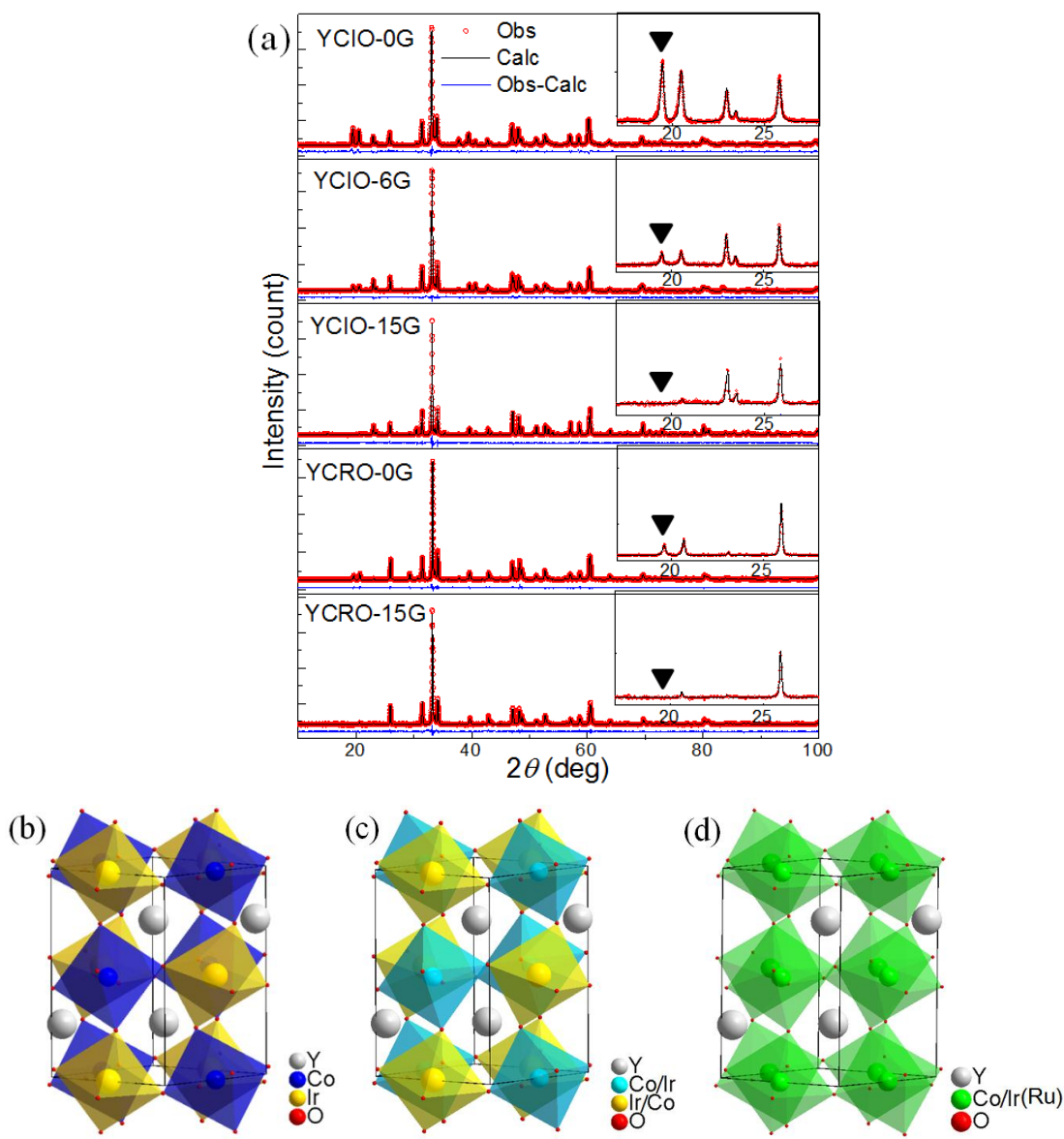

Figure 1 (a) From top to bottom: Room-temperature Rietveld refined PXD profiles of YCIO-0G, YCIO-6G, YCIO-15G, YCRO-0G and YCRO-15G. Insets show peaks around 19.5 degree, and black triangles mark the positions of (011) peaks. The crystal structures of (b) YCIO-0G, (c) YCIO-6G and (d) YCIO-15G and YCRO-15G. Note that $B'$-$B''$ antisite ratio is only 7.3% in YCIO-0G, while the value increases to 27.3% in YCIO-6G and 50% in YCIO-15G/YCRO-15G.

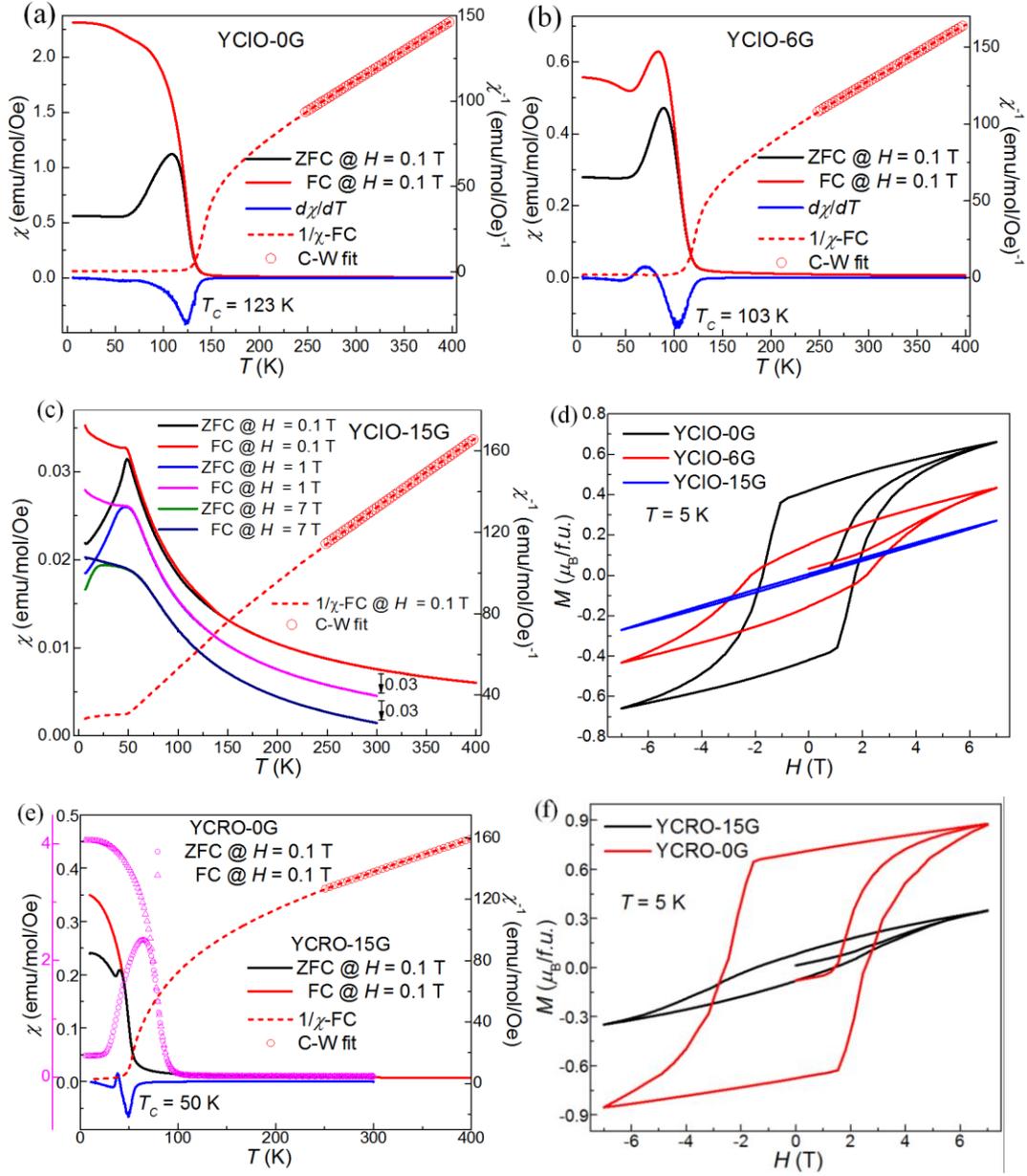

Figure 2 Magnetic properties of $Y_2CoIrO_6$ symthesized under varous prssures. Temperature dependence of magnetization ($M(T)$) and inverse susceptibility ($\chi^{-1}(T)$) at $H$ = 0.1 T for (a) YCIO-0G and (b) YCIO-6G. (c) $M(T)$ at varying fields and $\chi^{-1}(T)$ at $H$ = 0.1 T for YCIO-15G. (d) Field dependence of magnetization ($M(H)$) measured at 5 K for YCIO-0G, YCIO-6G and YCIO-15G. (e) $M(T)$ and $\chi^{-1}(T)$ at $H$ = 0.1 T for YCRO-0G and YCRO-15G. Note that YCRO-0G corresponds to purple *y*-axis and the black one is for YCRO-15G. (f) $M(H)$ measured at 5 K for YCRO-0G and YCRO-15G.

**Figure 3**

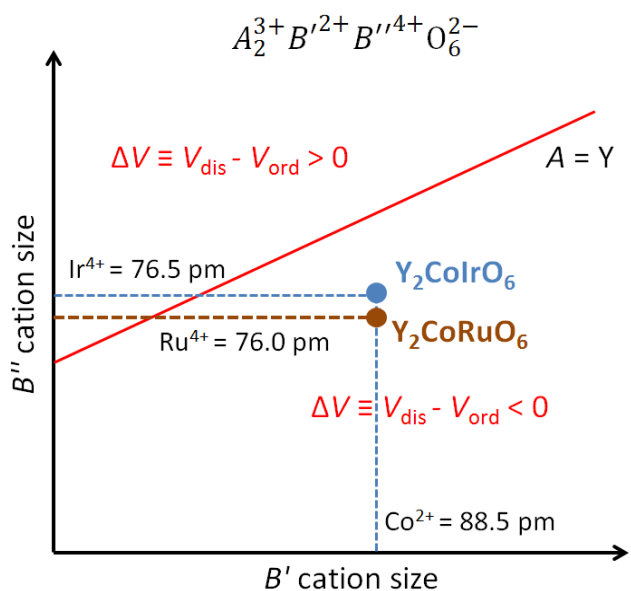

Figure 3 Systematic diagram for $A_2^{3+}B'^{2+}B''^{4+}O_6$ double perovskites. There are two regions: $\Delta V > 0$ and $\Delta V < 0$, which indicate that ordered and disordered phases are preferred upon pressure, respectively. The red line is a boundary of the two regions. Since the size of cation $Co^{2+}$ is substantially larger than that of $Ir^{4+}(Ru^{4+})$, $Y_2CoIrO_6$ and $Y_2CoRuO_6$ are in the region of $\Delta V < 0$.

**Figure 4**

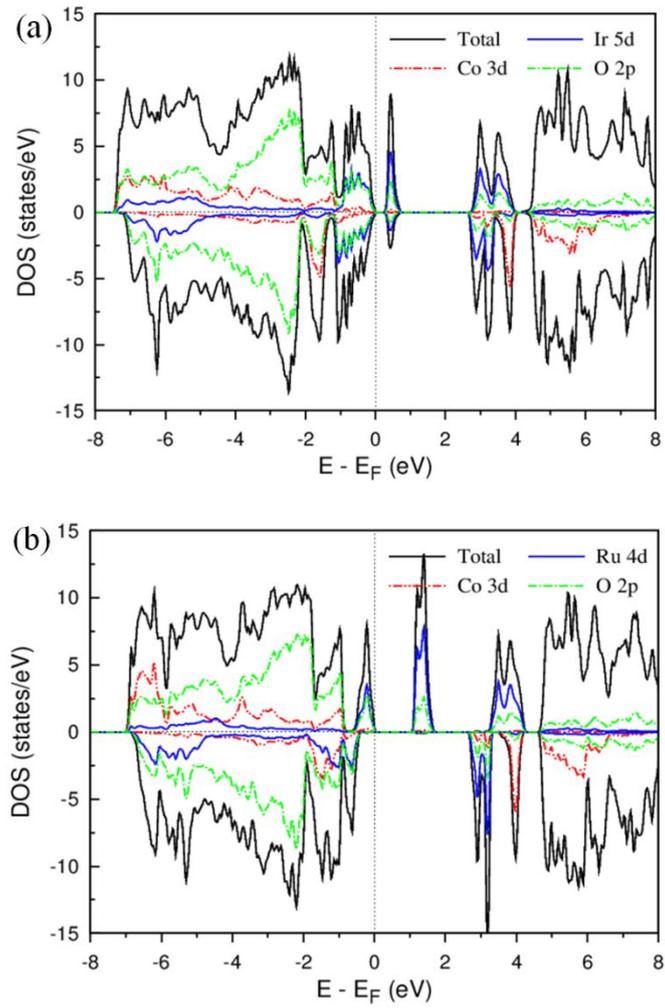

Figure 4 Total and partial density of states (DOS) of (a) $Y_2CoIrO_6$ and (b) $Y_2CoRuO_6$ with ferrimagnetic order from GGA+SOC+U calculations. The Fermi level is set to zero in the plot. The positive and negative values in DOS correspond to spin up and down, respectively.

**Table 1**

Table 1 Selected structural parameters as determined by Rietveld refinements.

| Parameters | YCIO-0G | YCIO-6G | YCIO-15G | YCRO-0G | YCRO-15G |
|---|---|---|---|---|---|
| synthesis pressure (GPa) | 0 | 6 | 15 | 0 | 15 |
| space group | $P2_1/n$ | $P2_1/n$ | $Pbnm$ | $P2_1/n$ | $Pbnm$ |
| $a$ (Å) | 5.2690(1) | 5.2585(1) | 5.2515(1) | 5.2683(1) | 5.2448(1) |
| $b$ (Å) | 5.6910(1) | 5.6902(1) | 5.6856(1) | 5.7073(1) | 5.6878(2) |
| $c$ (Å) | 7.5720(1) | 7.5621(2) | 7.5589(1) | 7.5561(1) | 7.5461(2) |
| $\beta$ (degree) | 90.099(2) | 89.997(4) | 90 | 89.942(2) | 90 |
| $V$ (Å$^3$) | 227.053(7) | 226.268(12) | 225.696(4) | 227.198(6) | 225.112(14) |
| $R_{wp}$ (%), $R_p$ (%) | 2.06, 1.52 | 2.62, 2.04 | 2.38, 1.71 | 2.54, 1.86 | 2.18, 1.69 |
| $B'$-$B''$ antisite (%), $s$ | 7.3, 0.854 | 27.3, 0.454 | 50, 0 | 8.0, 0.840 | 50, 0 |
| BVS-$B'$ | Co$^{2.12+}$ | N/A | N/A | Co$^{2.05+}$ | N/A |
| BVS-$B''$ | Ir$^{4.09+}$ | N/A | N/A | Ru$^{3.98+}$ | N/A |